\documentclass[prb,twocolumn,showpacs,amsmath,amssymb]{revtex4}
\usepackage{graphicx}
\usepackage{color}
\usepackage[active]{srcltx}

\newtheorem{teo}{Theorem}
\newtheorem{defi}{Definition}
\newtheorem{obs}{Remark}

\newcommand{\RR}{\mathbb R}
\newcommand{\CC}{\mathbb C}
\newcommand{\EE}{\mathbb E}
\newcommand{\NN}{\mathbb N}
\newcommand{\ZZ}{\mathbb Z}
\newcommand{\al }{\alpha }
\newcommand{\Ga }{\Gamma }
\newcommand{\e }{\epsilon }
\newcommand{\be }{\beta }
\newcommand{\TT}{\mathbb T}

\begin{document}

\title{Infinite ergodic theory and Non-extensive entropies.}
\author{L.M. GAGGERO-SAGER, E.R. PUJALS AND O. SOTOLONGO-COSTA}
\affiliation{$^1$ Universidad Aut\'onoma del Estado de Morelos. Facultad de Ciencias. Ave. Universidad 1001, Cuernavaca, Morelos, M\'exico\\
$^2$ Instituto de Matem\'atica Pura e Aplicada - IMPA. Dona Castorina 110, Rio de Janeiro. Brasil \\
$^3$ Facultad de Fisica, Universidad de la Habana. La Habana, Cuba. }

\begin{abstract}
We bring into account a series of result in the infinite ergodic theory that
we believe that they are relevant to the theory of non-extensive entropies.
\end{abstract}

\maketitle

\affiliation{$^1$ Universidad Aut\'onoma del Estado de Morelos. Facultad de Ciencias. Ave. Universidad 1001, Cuernavaca, Morelos, M\'exico.\\
$^2$ Instituto de Matem\'atica Pura e Aplicada - IMPA. Dona Castorina 110, Rio de Janeiro. Brasil. \\
$^3$ C\'atedra "Henri Poncar\'e" de Sistemas Complejos, Universidad de la Habana. La Habana, Cuba. }

\section{Introduction.}

\label{intro}

In Ref. [\onlinecite{Tsallis}] it has been proposed that Non-extensive
entropies can be useful to describe the dynamics of systems with zero
Lyapunov exponent but exhibiting some weak form of sensitiveness to initial
conditions. This sensitiveness would be described by the $q-$generalized
Lyapunov exponent $\lambda _{q}$ for some value of $q$ in such a way that
the average distance between point after $n-$iterates would be of the order
of $\exp _{q}(\lambda _{q}n)$ where $\exp _{q}(t)=[1+(1-q)t]^{\frac{1}{1-q}}$%
. In particular, it was conjectured in Ref. [\onlinecite{Tsallis2}] that a
version of Pesin's theorem for sub-exponential instability would relate the $%
q-$entropy with the $q-$generalized Lyapunov exponent. More precisely, they
would coincide if $\lambda _{q}>0$ and $q<1$ and in this case the average
distance increases polynomially.

With this in mind, we are recalling some results in the infinite ergodic
theory, meaning the ergodic theory of systems preserving a non-finite
measure. The reason to focus on these type of systems is because they
exhibit zero Lyapunov exponent and they may exhibit sub-exponential
instability. Therefore, they can be analyzed in the framework of
non-extensive entropies. 

In this direction, our goal is to show that for some systems preserving an
infinite measure (and therefore having zero Lyapunov exponent) the next
assertions  hold:

\begin{enumerate}
\item  they do not have a unique quantity that describes sub-exponential
instabilities;

\item  sub-exponential rates can grow faster than polynomial ones.
\end{enumerate}

As a consequence of that, a twofold observation follows: \emph{there is no a single quantity
to characterize sub-exponential growth, and it can not necessary  be understood   in polynomial terms.}

To explain that, the following is done:

\begin{enumerate}
\item   A class of examples having infinite invariant measure  is introduced;

\item It is explained why systems  having infinite invariant measure exhibit zero Lyapunov exponents;

\item  It is shown how ergodic theorems can be recovered for infinite
invariant measure and how the time averages become intrinsically random
making difficult to find a unique ``generalized Lyapunov'' quantity;

\item  It is explained how these type of systems displays sub-exponential
instabilities and it is shown   different examples where the sub-exponential instability is not
polynomial.

\end{enumerate}

For this exposition, we follow Ref. [\onlinecite{Aaronson}] and Ref. 
[\onlinecite{Zew}]. We want to point out that in this note, no new theorems
for the infinite ergodic theory are provided; we only present some results
already proved somewhere else and we recast them to show that they could be relevant for the theory of 
non-extensive statistical mechanics.

\section{Infinite measure.}

\label{infinite}

Given a map $T:X\rightarrow X$ acting on a phase space $X$, its action can
lead to very complicated (chaotic) dynamics. Ergodic theory can be seen as a
quantitative theory of dynamical systems, enabling us to rigorously deal
with such situations, where it is impossible to predict when exactly some
relevant event is going to take place. For example, Birkhoff ergodic
theorem, tells us quite precisely how often an event will occur for typical
initial states. In fact, a  rich quantitative theory is available for systems
possessing an invariant finite measure $\mu $, meaning that $\mu \circ
T^{-1}=\mu $. Moreover, in case of smooth systems, Birkhoff ergodic theorem
allows to characterize the rate of mixing of a system.

However, there do exist systems of interest (not necessarily too exotic),
which happen to have an infinite invariant measure, i.e., measure preserved by $T$  and that 
$\mu (X)=\infty $. The
``Infinite Ergodic Theory'' focuses on such systems trying to answer the
simplest quantitative question of understanding the long-term behavior of
occupation times 
\begin{eqnarray}
 S_{n}(A):=\Sigma _{k=0}^{n-1}1_{A}(T^{k}(x))
\end{eqnarray}
where $1_{A}$ is the characteristic function of the set $A$ ($1_A(x)=1$ if $x\in A$ otherwise the value is zero).
The  quantity  $S_n(A)$ counts the number of visits an orbit pays to $A$
before time $n.$ Slightly more general, we can also look at ergodic sums 
\begin{eqnarray}
 S_{n}(f):=\Sigma _{k=0}^{n-1}f(T^{k}(x))
\end{eqnarray}
of measurable functions $f$. 

Some typical examples of infinite measure preserving transformation are:
\begin{enumerate}
\item  Boole maps, $T:\RR \setminus \{0\}\rightarrow \RR $, $T(x)=x-\frac{1}{x}$,
where the invariant measure is the Lebesgue measure in the Real line (see Ref. 
[\onlinecite{B}] and [\onlinecite{AW}]);

\item  Pomeau-Manneville maps, $T:[0,1]\rightarrow \lbrack 0,1]$, $T(x)=x+cx^{p}mod(1)$, 
in which zero is a parabolic fixed point ($T^{\prime
}(0)=1$), and the invariant measure has support in $[0,1]$ but gives 
infinite measure to the interval (see Ref. [\onlinecite{PM}, \onlinecite{CI}]);

\item Polynomial and  rational maps on $\CC$ (quotient of polynomials acting on $\CC$) with parabolic 
fixed points (points where the derivative has modulus one) in the Julia set and no critical 
points there, where the invariant measure is a $h-$conformal measure concentrated in the Julia set 
and $h$ is the Hausdorff dimension of the Julia set  (see Ref. [\onlinecite{ADU}]);

\item  Some quadratic unimodal maps (or logistic type maps) 
where the invariant measure is absolute
continuous and giving infinite measure to the domain  (see Ref. [\onlinecite{HK}, 
\onlinecite{Br}, \onlinecite{ABJ}]);

\item  Horocycle flows on infinite regular covers of compact hyperbolic surfaces, where the invariant
measure is the classical volume measure (see Ref. [\onlinecite{Ledra1}]);

\item  Two dimensional version of the Boole's map, $T:\RR^{2}\setminus
\{0\}\times \RR \rightarrow \RR^{2}$, $T(x,y)=(x-\frac{1}{y},x+y-\frac{1}{y})$,
where the invariant measure is the Lebesgue measure in the whole two dimensional plane
(see Ref. [\onlinecite{Devaney}]);

\item  Special flows (see Ref. [\onlinecite{O}] and [\onlinecite{SK}]) which are
volume preserving flows of the two dimensional torus $\TT^2$ given by solution of
the equation 
$$\begin{array}{ccccc}
\dot{x}=f(x,y)  \\ 
\dot{y}=f(x,y)\al 
\end{array}$$
where $\al$ is irrational, and  $f:\TT^{2}\rightarrow \RR$ verifies that $f(0,0)=0$ and $
f(x,y)>0\,\forall \,(x,y)\neq (0,0).$
\end{enumerate}

The first two systems are conjugated to the classical doubling function
acting on the interval $[0,1]$ and so their dynamics can be described by the
symbolic shift acting on the space of sequences of two symbols. In
particular, they are topologically mixing and have infinitely many periodic
orbits. The Pomeau-Manneville maps were introduced to
model intermittent behavior in fluid dynamics (see Ref. [\onlinecite{PM}]). 
For those type of system, the resulting behavior
is an alternation of chaotic (when the orbit stays far from the parabolic point and where
 the map is similar to the baker’s map) and regular
when the orbit is trapped near the parabolic point.  On the other hand, the presence of a parabolic
fixed point makes those systems to have zero Lyapunov exponent and to
display an infinite invariant measure.
Moreover,  for those type of systems it had been calculated  rigorously the information content (Kolmogorov
complexity) of the symbolic orbits generated by these systems (see Ref. [\onlinecite{BG}, \onlinecite{G}]),
showing a behavior for the information
that is between a positive entropy system (the information
grows linearly with time) and an integrable system (the information grows as the logarithm of time).

Polynomial and rational maps are the typical dynamics acting on the complex plane; the classical example is given by
the family of quadratical polynomials $P_\mu(z)=z^2+\mu$. Recall that the Julia set (in the case of polynomials) is defined as the boundary 
of the set of points that its trajectory does not escape to infinite and its concentrate all the dynamics complexity (see for instance   Ref. [\onlinecite{Milnor}]).

The quadratic family is the classical example of one-dimensional real dynamics exhibiting chaotic dynamics.
In Ref. [\onlinecite{HK}, \onlinecite{Br}, \onlinecite{ABJ}] is shown that for certain parameters, the associated map has an infinite absolute continuous  invariant measure.
 
The two dimensional version of the Boole's map introduced by Henon  to model certain problem in celestial mechanics (see Ref. [\onlinecite{Henon1}, \onlinecite{Henon2}), are conjugated to the Baker map acting on a two
dimensional rectangle (see Ref. \onlinecite{Devaney}]).

The Horocyclic flow on compact hyperbolic surfaces is the most 
classical example of minimal and ergodic dynamic respect to finite volume measure  (fact proved by Hedlund in 1930). 
They are associated to the classical geodesic flow on hyperbolic surfaces (free motion on hyperbolic surfaces). 
However, as proved in  Ref. [\onlinecite{Ledra1}], when is considered infinite covers, the measure become infinite.

The last example, is a topologically mixing conservative system on the torus and having only one periodic point (the point $(0,0)$ which is fixed). Those systems are semiconjugated to quasi-periodic ones. The fact that the fixed point is parabolic, allows to find an infinite invariant measure. 

Some of those systems were treated in the context of non-extensive
entropies, see for instance  Ref. [\onlinecite{TPZ}, \onlinecite{CLPT}] for the
case of the quadratic family. We want to point out that there are many other 
systems having zero Lyapunov exponent but exhibiting weak form of mixing and which are not covered by the above list (see for instance  Ref. [\onlinecite{F1}, \onlinecite{F1}, \onlinecite{FH}, \onlinecite{AK}, \onlinecite{FS}, \onlinecite{KK}]). However, in some cases it is unknown if they have infinite invariant measure

Even though many of the above described systems  could
seem very restrictive, it is important to remark that for one-dimensional
dynamics, whenever a parabolic periodic point appear, they present anomalous statistical
behavior and for this reason they have been used
as models of many interesting physical systems.

The first ergodic theorem for recurrent   ergodic measure transformation (i.e., almost every point is recurrent and any invariant set or its complement has measure zero) in the context of infinite measure is the
following:
\begin{teo}
\label{Birkhoff thm} (Birkhoff's Pointwise Ergodic Theorem). Let  $T$ be 
a recurrent  ergodic measure transformation on the infinite measure space $(X;A;\mu),$ 
then 
\begin{eqnarray}
 \frac{1}{n}S_{n}(f)\rightarrow 0.
\end{eqnarray}
\end{teo}

This theorem shows that smooth systems preserving an infinite measure has 
\emph{zero Lyapunov exponent}. More precisely:
\begin{obs}
\label{Birkhoff obs} Observe that if $T$ is a smooth one dimensional map on
the line and $\log {T^{\prime }}\in L^{1}(\mu )$ then it follows that 
\begin{eqnarray}
 \frac{1}{n}\log {{T^{n}}^{\prime }(x)}=\frac{1}{n}S_{n}(\log {T^{\prime }(x)})\rightarrow 0,\,\,\,x\,\mbox{a.e}
\end{eqnarray}
meaning that for almost every point the Lyapunov exponent is zero.
\end{obs}
It is natural to ask if it is possible to find a sequence $\{a_{n}\}$ of
positive normalizing constants, such that for all $A\in \mathcal{A}$,
follows that $\frac{1}{a_{n}}S_{n}(1_{A})\rightarrow \mu (A)$ and for any 
$f\in L^{1}(\mu )$ follows that $\frac{1}{a_{n}}S_{n}(f)\rightarrow \int
fd\mu $. This could  be regarded as an appropriate version of the
ergodic theorem for spaces with infinite measure. The following shows that
this is not possible:

\begin{teo}
Let $T$ be a recurrent ergodic measure transformation on the infinite measure space $(X;A;\mu),$ 
 and let any sequence $\{a_{n}\}_{n\in }$ . Then for
all $f\in L^{1}(\mu )$ either 
\begin{eqnarray}
 \limsup \frac{1}{a_{n}}S_{n}(f)=0
\end{eqnarray}
or
\begin{eqnarray}
 \limsup \frac{1}{a_{n}}S_{n}(f)=\infty.
\end{eqnarray}

\end{teo}

The previous theorem shows that any potential normalizing sequence either
over or underestimates the actual size of ergodic sums. Moreover, this
 points out that in infinite measure systems, the time average of
an observation function fluctuates.

To find the appropriate normalizing sequences $a_{n}$ (or time rescaling),
it is needed to define the dual operator $\hat{T}:L^{1}(\mu )\rightarrow
L^{1}(\mu )$ given by $\hat{T}(f)=f\circ T^{-1},$ which describes the
evolution of measures under the action of $T$ in the level of densities:

\begin{defi}
It is said that the system is pointwise dual ergodic, if there exist
constants $a_{n}=a_{n}(T)$, $n\in \NN$, such that for any $L^{1}(\mu )$
it follows that 
\begin{eqnarray}
 \lim \frac{1}{a_{n}}\Sigma _{k=0}^{n-1}\hat{T}^{k}(f)=\int f\,du
\end{eqnarray}
\end{defi}

The sequence $a_{n}(T)$ is uniquely determined up to asymptotic equality,
and is called the return sequence of $T.$ Moreover, when the map is
pointwise dual ergodic, there exists sets $A\in \mathcal{A}$ with $\mu
(A)<+\infty $ such that 
\begin{eqnarray}
 \lim \frac{1}{a_{n}}\Sigma _{k=0}^{n-1}\hat{T}^{k}(1_{A})=\mu (A).
\end{eqnarray}
This type of sets are called \emph{Darling-Kac (DK) sets}. This means that
the measure of the set $A$ is recovered by pulling back the Lebesgue measure
in $A$ and averaged by the sequences $\{a_{n}\}$. The  next result shows that the proper normalized time rescaling of an observation function converges in distribution, provided that the return sequence has certain properties.

If the return sequence is regularly varying with
index $\al\in \lbrack 0,1]$ (i.e. $a_{n}(T)=n^{\al}h(n)$ and for any $c>0$ 
$\frac{h(cn)}{h(n)}\rightarrow c^{\al}$), the
asymptotic behavior of $S_{n}(f)$ can be described almost surely in
distribution as follows, whenever it is assumed that the map is a recurrent
ergodic measure transformation:

\begin{teo}
\label{ADK thm} (Aaronson's Darling-Kac Theorem) Let T be a recurrent
ergodic measure transformation on the infinite measure space $(X;A;\mu )$.
Assume there is some DK-set $A\in \mathcal{A}$. If $a_{n}=a_{n}(T)$ is
regularly varying of index $1-\alpha $ (for some $\al\in \lbrack 0;1]$),
then for all $f\in L^{1}(\mu )$ and all $t>0$ 
\begin{eqnarray}
 \mu _{A}(\frac{1}{a_{n}}S_{n}(f)<t)\rightarrow Pr[M_{\al}(t)\int_{X}fd\mu
]\,\,\,\mbox{as}\,\,n\rightarrow \infty.
\end{eqnarray}
\end{teo}

In previous theorem, $\mu _{A}$ denotes the measure $\mu $ restricted to the
set $A$ (and can be replaced by any probability absolutely continuous
respect to $\mu $), and $M_{\al},\al\in \lbrack 0;1]$ denotes a non-negative
real random variable distributed according to the (normalized)
Mittag-Leffler distribution of order $\al$, which can be characterized by
its moments
\begin{eqnarray}
 \EE[M^r_\al]= r!\frac{\Ga(1+\al)^r}{\Ga(1+r\al)},\,\, r\geq 0.
\end{eqnarray}
Going back to the problem of finding the sub-exponen\-tial Lyapunov exponents,
the previous theorem shows that in certain cases it is not possible to find
a unique quantity for almost every point even in the case of ergodic
systems.  In fact, assuming that the transformation $T$ is one-dimensional and smooth,
 if $f=\log (T^{\prime })$ then  the hypothesis of theorem \ref
{ADK thm} hold and $\al\neq 0$, it follows that given three points 
$t_{1}<t_{2}<t_{3}$ there is a set of positive measure of initial conditions
such that $\exp (a_{n}t_{1})<(T^{n})^{\prime }(x)<\exp (a_{n}t_{2})$
(provided $n$ large) and a set of positive measure of initial conditions
such that $\exp (a_{n}t_{2})<(T^{n})^{\prime }(x)<\exp (a_{n}t_{3}).$ 

\bigskip 

In particular$,$ \emph{it would not be possible to find a simple Pesin's
type formula that relates the ``sub-exponential Lyapunov exponent'' with a
unique quantity as the generalized entropy}.

\bigskip 

However, theorem \ref{ADK thm} gives the range of fluctuation of the
Lyapunov exponent and the sequences $a_{n}$ provides up to a constant that
depends on set of initial conditions, the rate of separation of
trajectories. In fact for almost every point $x$, there exists $t(x)$ such
that for any $\e>0$ if $n$ is large enough then 
\begin{eqnarray}
 \exp (a_{n}(t(x)-\e))<(T^{n})^{\prime }(x)<\exp (a_{n}(t(x)+\e)).
\end{eqnarray}
So, for some of the maps described before, we are  going to explicit  the normalizing  sequences $a_{n}=a_{n}(T)$  and we are going to apply theorem \ref{ADK thm}:

\begin{enumerate}
\item  For the Pomeau-Mannivelle maps, $a_{n}=n^{\frac{1}{p}}$ if $p>1$, 
$a_{n}=\frac{n}{\log {n}}$ if $p=1$;

\item  For the Boole map, $a_{n}=\sqrt{n};$

\item  For rational maps on on $\CC$ with parabolic points in the Julia set, $a_n=n^{\be-1}$ for $1<\be<4$ and $a_n= \frac{n}{\log{n}}$ if $\be=4$ with  $\be=\frac{p}{p+1}h$ where $p$ is the first integer larger than one such that 
the derivative at the parabolic fixed point does not vanish and $h$ is the Hausdorff dimension of the Julia set $J$;

\item  Horocycle flows on periodic hyperbolic surfaces, $a(t)=\frac{t}{\ln
(t)^{k}}$ with $k\in \frac{1}{2}$ depending on the surface (see Ref. [\onlinecite{Ledra2}]).
\end{enumerate}

Observe that in same cases, the rescaling of time has a polynomial fashion,
but the situation for the asymptotic growth of the derivative is quite
different. So now, using that in the one-dimensional $\log({T^n}
^{\prime}(x))=\Sigma_{j=0}^{n-1} \log(T^{\prime}(T^j(x)))$, theorem \ref{ADK
thm} and the explicit calculation of the sequences $a_n$ one can provide the
asymptotic growth of the derivative and therefore the rate of
sub-exponential instability:

\begin{enumerate}
\item  \emph{Sub-exponential Lyapunov exponents for Pomeau-Manneville maps:}
Given any pair of positive numbers $t_{1}<t_{2}$, in the case that $p>1$ it
follows that for large $n$ there is a positive set of initial condition that 
$\exp (t_{1}n^{\frac{1}{p}})<(T^{n})^{\prime }(x)<\exp (t_{2}n^{\frac{1}{p}})
$. If $p=1$ it follows that for large $n$ there is a positive set of initial
condition that $\exp (t_{1}\frac{n}{\ln (n)})<(T^{n})^{\prime }(x)<\exp
(t_{2}\frac{n}{\ln (n)})$.

\item {\em Sub-exponential Lyapunov exponents for maps on $\CC$ with
parabolic points in the Julia set:} it is similar to the Pomeau-Manneville
maps.

\item  \emph{Sub-exponential Lyapunov exponents for $\,$ Boole maps:} There
is a positive set of initial conditions for which $\exp (t_{1}\sqrt{n})<(T^{n})^{\prime }(x)<\exp (t_{2}\sqrt{n})$.
\end{enumerate}

In any case, it follows that \emph{the sub-exponential growth of the
derivative by iteration is larger than the growth of any polynomial and
therefore it can not be described by any $\exp _{q}$  for any value of $q.$}

For the case of quadratic family, no explicit calculation have been
performed, however a vast range of different type of sub-instability can be
expected. This is discussed latter at the end of the present section.

Even theorem \ref{ADK thm} shows that normalized time averages only converge
to distributions, it is natural to wonder if the double average of the
weighted Birkhoff sums converge. In fact, the following theorem shows how
the expected value of normalized time averages also has a limit: 

\begin{teo}
Assume that $a_{n}(T)=n^{\al}h(n)$ is regularly varying with index $\al\in
(0,1]$ or $\al=0$ and with $h(n)\approx \exp (\int_{1}^{t}\frac{\eta (t)}{t}
dt)$, where $\eta $ is monotonic, $\eta (t)\rightarrow 0$ as $t\rightarrow
\infty $ and $\frac{\eta (t)}{\log {t}}\rightarrow \infty $ as $t\rightarrow
\infty $. Then for any function $f\in L^{1}(u)$ 
\begin{eqnarray}
 \lim \frac{1}{N}\Sigma \frac{1}{a_{n}(T)}S_{n}(f)=\int f\,du
\end{eqnarray}
in measure. 
\end{teo}

It is important also to point out, that in certain cases (for instance for
the Boole maps and Pomeau-Mannivelle maps), the rates $a_{n}(T)$ are related
to the {\em induced map (or return map)}: Given a set $A$ with $\mu (A)<+\infty $ and assuming that almost every
return point (which is the case in the examples considered and in the
hypothesis of the theorems), then for almost every point $x$ it can be
defined $n(x)=\min \{n\geq 1:T^{n}(x)\in A\}$ and then, is defined the map $
x\rightarrow T^{n(x)}.$ It turns out that the measure $\mu $ restricted to $A
$ is ergodic and finite. Moreover, the examples of unimodal maps
(quadratic-type maps) with infinite measure is obtained through a return map
construction (usually called towers) and showing a type of non-integrability
condition for the return times (see  Ref. [\onlinecite{ABJ}] for details). These
analysis would provide a precise description of the sub-exponential
instability. Precise studies  of  quantitative recurrence in systems having an infinite invariant
measure have been performed also in Ref. [\onlinecite{GKP}]. Moreover, in Ref. [\onlinecite{BGI}]
is done, through a series of examples, a detailed investigation of the relationships
between quantitative recurrence indicators and algorithmic complexity of orbits
in weakly chaotic dynamical systems.

\end{document}